\documentclass[a4paper]{article}
\usepackage{amsmath}
\usepackage{amssymb}

\newtheorem{theorem}{Theorem}

\newtheorem{definition}[theorem]{Definition}

\newtheorem{proposition}[theorem]{Proposition}

\begin{document}

\title{Generalizing the autonomous Kepler Ermakov system in a Riemannian
space}
\date{}
\author{Michael Tsamparlis\thanks{%
Email: mtsampa@phys.uoa.gr} \ and Andronikos Paliathanasis\thanks{%
Email: anpaliat@phys.uoa.gr} \\
{\small \textit{Faculty of Physics, Department of Astrophysics - Astronomy -
Mechanics,}}\\
{\small \textit{\ University of Athens, Panepistemiopolis, Athens 157 83,
GREECE}}}
\maketitle

\begin{abstract}
We generalize the two dimensional autonomous Hamiltonian Kepler Ermakov
dynamical system to three dimensions using the $sl(2,R)$ invariance of
Noether symmetries and determine all three dimensional autonomous
Hamiltonian Kepler Ermakov dynamical systems which are Liouville integrable
via Noether symmetries. Subsequently we generalize the autonomous Kepler
Ermakov system in a Riemannian space which admits a gradient homothetic
vector by the requirements (a)\ that it admits a first integral (the
Riemannian Ermakov invariant) and (b) it has $sl(2,R)$ invariance. We
consider both the non-Hamiltonian and the Hamiltonian systems. In each case
we compute the Riemannian Ermakov invariant and the equations defining the
dynamical system. We apply the results in General Relativity and determine
the autonomous Hamiltonian Riemannian Kepler Ermakov system in the spatially
flat Friedman Robertson Walker spacetime. We consider a locally rotational
symmetric (LRS) spacetime of class A and discuss two cosmological models.
The first cosmological model consists of a scalar field with exponential
potential and a perfect fluid with a stiff equation of state. The second
cosmological model is the $f(R)$ modified gravity model of $\Lambda _{bc}$CDM%
$.$ It is shown that in both applications the gravitational field equations
reduce to those of the generalized autonomous Riemannian Kepler Ermakov
dynamical system which is Liouville integrable via Noether integrals.
\end{abstract}


PACS - numbers: 45.20.D-, 02.20.Sv, 02.40.Dr, 02.40.Ky, 02.30.Ik, 98.80.-k

\section{Introduction}

The Ermakov system has its roots in the study of the one dimensional time
dependent harmonic oscillator
\begin{equation}
\ddot{x}+\omega ^{2}(t)x=0.  \label{AEP.00.1}
\end{equation}%
Ermakov \cite{Ermakov} obtained a first integral $J$ of this equation by
introducing the auxiliary equation
\begin{equation}
\ddot{\rho}+\omega ^{2}(t)\rho =\rho ^{-3}  \label{AEP.00.2}
\end{equation}%
eliminating the $\omega ^{2}(t)$ term and multiplying with the integrating
factor $\rho \dot{x}-\dot{\rho}x:$%
\begin{equation}
J=\frac{1}{2}\left[ (\rho \dot{x}-\dot{\rho}x)^{2}+(x/\rho )^{2}\right] .
\label{AEP.00.3}
\end{equation}

The Ermakov system was rediscovered nearly a century after its introduction
\cite{Lewis (1967)} and subsequently was generalized beyond the harmonic
oscillator to a two dimensional dynamical system which admits a first
integral \cite{Ray Reid}. In a series of papers the Lie, the Noether and the
dynamical symmetries of this generalized system have been studied. A short
review of these studies \ and a detailed list of relevant references can be
found in \cite{Leach Andriopoulos}. Earlier reviews of the Ermakov system
and its numerous applications in divertive areas of Physics can be found in
\cite{Rogers et all,Schief et all}.

The general Ermakov system does not admit Lie point symmetries. The form of
the most general Ermakov system which admits Lie point symmetries has been
determined in \cite{Goedert Haas 1998} and it is called the Kepler Ermakov
system \cite{Athorne 1991,Leach1991}. It is well known that these Lie
symmetries are a representation of the $sl(2,R)$ algebra.

In an attempt to generalize the Kepler Ermakov system to higher dimensions
Leach \cite{Leach1991} used a transformation to remove the time dependent
frequency term and then demanded that the autonomous `generalized' Kepler
Ermakov system will posses two properties (a)\ A first integral, the Ermakov
invariant and (b) $sl(2,R)$ invariance to Lie symmetries. It has been shown
that the invariance group of the Ermakov invariant is reacher than $sl(2,R)$%
\ \cite{Govinder Leach 1994}. The purpose of the present work is to use
Leach's proposal and generalize the autonomous Kepler Ermakov system in two
directions (a) to higher dimensions using the $sl(2,R)$ invariance with
respect to Noether symmetries (provided the system is Hamiltonian) and (b)
in a Riemannian space which admits a gradient homothetic vector (HV).

The generalization of the autonomous Kepler Ermakov system to three
dimensions using Lie symmetries has been done in \cite{Leach1991}. In the
following we use the results of \cite{DynamicalS} to generalize the subset
of autonomous Hamiltonian Kepler Ermakov systems to three dimensions via
Noether symmetries. We show that there is a family of three dimensional
autonomous Hamiltonian Kepler Ermakov systems parametrized by an arbitrary
function $f$ which admits the elements of $sl(2,R)$ \ as Noether
symmetries.\ Each member of this family admits two first integrals, the
Hamiltonian and the Ermakov invariant.

We use this result in order to determine all three dimensional Hamiltonian
Kepler Ermakov systems which are Liouville integrable via Noether
symmetries. To do this we need to determine all members of the family, that
is those functions $f,$ for which the corresponding system admits an
additional Noether symmetry.

The results of \cite{DynamicalS} indicate that there are two cases to be
considered i.e. Noether point symmetries resulting from linear combinations
of (a) translations and (b) rotations (elements of the $so\left( 3\right) $
algebra). In each case we\ determine the functions $f$ and the required
extra time independent first integral.

The above scenario can be generalized to an $n$- dimensional Euclidian space
as Leach indicates in \cite{Leach1991}, however at the cost of major
complexity and number of cases to be considered. Indeed as it can be seen by
the results of \cite{DynamicalS}, the situation is complex enough even for
the three dimensional case .

We continue with the generalization of the Kepler Ermakov system in a
different and more drastic direction. We note that the Ermakov systems
considered so far are based on the Euclidian space, therefore we may call
them \emph{Euclidian} Ermakov systems. Furthermore the $sl(2,R)$ symmetry
algebra of the autonomous Kepler Ermakov system is generated by the trivial
symmetry $\partial _{t}$ and the gradient HV\ of the Euclidian two
dimensional space $E^{2}.$ Using this observation we generalize the
autonomous Kepler Ermakov system (not necessarily Hamiltonian) in a $n-$%
dimensional Riemannian space which admits a gradient HV using either Lie or
Noether symmetries. The new dynamical system we call the Riemannian Kepler
Ermakov system. This generalization makes possible the application of the
autonomous Kepler Ermakov system in General Relativity and in particular in
Cosmology.

Concerning General Relativity we determine the four dimensional autonomous
Riemannian Kepler Ermakov system and the associated Riemannian Ermakov
invariant in the spatially flat Freedman - Robertson - Walker (FRW)\
spacetime; we use the results of \cite{DynamicalS,Tsam10} to calculate the
extra Noether point symmetries. The applications to cosmology concern two
models for dark energy on a locally rotational symmetric (LRS) space time.
The first model involves a scalar field with an exponential potential
minimally interacting with a perfect fluid with a stiff equation of state.
The second cosmological model is the $f(R)$ modified gravity model of $%
\Lambda _{bc}CDM$. It is shown that in both models the gravitational field
equations define an autonomous Riemannian Kepler Ermakov system which is
integrable via Noether integrals.

The structure of the paper is as follows. In section \ref{ErmakovSys} we
review the main features of the two dimensional autonomous Euclidian Kepler
Ermakov system. In section \ref{Generalizing the Kepler Ermakov system} we
discuss the general scheme of generalization of the two dimensional
autonomous Euclidian Kepler Ermakov system to higher dimensions and to a
Riemannian space which admits a gradient HV. In section \ref{The 3d
Euclidian Kepler Ermakov system} we consider the generalization to the 3d
autonomous Euclidian Hamiltonian Kepler Ermakov system by Noether symmetries
and determine all such systems which are Liouville integrable. In section %
\ref{The Riemannian Kepler Ermakov system} we define the autonomous
Riemannian Kepler Ermakov system by the requirements that it will admit (a)
a first integral (the Ermakov invariant) and (b) posses $sl(2,R)$
invariance. In section \ \ref{The non conservative Riemannian Kepler Ermakov
system} we consider the non-conservative autonomous Riemannian Kepler
Ermakov system and derive the Riemannian Ermakov invariant and in section %
\ref{The Hamiltonian Kepler Ermakov system in a n-dimensional Riemannian
space} we repeat the same for the autonomous Hamiltonian Riemannian Kepler
Ermakov system. In the remaining sections we discuss the applications of the
autonomous Hamiltonian Riemannian Kepler Ermakov system in General
Relativity and in Cosmology. Finally in section \ref{Conclusion} we draw our
conclusions.

\section{The two-dimensional autonomous Kepler Ermakov system}

\label{ErmakovSys}

In \cite{Goedert Haas 1998} Hass and Goedert considered the most general 2d
Newtonian Ermakov system to be defined by the equations:%
\begin{align}
\ddot{x}+\omega ^{2}(t,x,y,\dot{x},\dot{y})x& =\frac{1}{yx^{2}}f\left( \frac{%
y}{x}\right)  \label{EqnEr.50} \\
\ddot{y}+\omega ^{2}(t,x,y,\dot{x},\dot{y})y& =\frac{1}{xy^{2}}g\left( \frac{%
y}{x}\right) .  \label{EqnEr.51}
\end{align}%
This system admits the Ermakov first integral
\begin{equation}
I=\frac{1}{2}(x\dot{y}-y\dot{x})^{2}+\int^{y/x}f\left( \tau \right) d\tau
+\int^{y/x}g\left( \tau \right) d\tau .  \label{EqnEr.52}
\end{equation}%
If one considers the transformation:%
\begin{align*}
\Omega ^{2}& =\omega ^{2}-\frac{1}{xy^{3}}g\left( \frac{y}{x}\right) \\
F\left( \frac{y}{x}\right) & =f\left( \frac{y}{x}\right) -\frac{x^{2}}{y^{2}}%
g\left( \frac{y}{x}\right)
\end{align*}%
then equations (\ref{EqnEr.50}),(\ref{EqnEr.51}) take the form:%
\begin{align}
\ddot{x}+\Omega ^{2}(x,y,\dot{x},\dot{y})x& =\frac{1}{x^{2}y}F\left( \frac{y%
}{x}\right)  \label{EqnEr.53} \\
\ddot{y}+\Omega ^{2}(x,y,\dot{x},\dot{y})y& =0.  \label{EqnEr.54}
\end{align}

Due to the second equation, except for special cases, the new function $%
\Omega $ is independent of $t$ and depends only on the dynamical variables $%
x,y$ and possibly on their derivative. The Ermakov first integral in the new
variables is:%
\begin{equation}
I=\frac{1}{2}(x\dot{y}-y\dot{x})^{2}+\int^{y/x}F\left( \lambda \right)
d\lambda .  \label{EqnEr.55}
\end{equation}%
The system of equations (\ref{EqnEr.53}), (\ref{EqnEr.54}) defines the most
general 2d Ermakov system and produces all its known forms for special
choices of the function $\Omega .$ For example the weak Kepler Ermakov
system \cite{Leach1991} defined by the equations \cite{Athorne 1991}
\begin{eqnarray}
\ddot{x}+\omega ^{2}(t)x+\frac{x}{r^{3}}H\left( x,y\right) -\frac{1}{x^{3}}%
f\left( \frac{y}{x}\right) &=&0  \label{Kepler-Ermakov 01} \\
\ddot{y}+\omega ^{2}(t)y+\frac{y}{r^{3}}H\left( x,y\right) -\frac{1}{y^{3}}%
g\left( \frac{y}{x}\right) &=&0  \label{Kepler-Ermakov 02}
\end{eqnarray}%
where $H,f,g~$are arbitrary functions of their argument, is obtained by the
function%
\begin{equation}
\Omega ^{2}(x,y)=\omega ^{2}(t)+H\left( x,y\right) /r^{3}  \label{EqnEr.55d}
\end{equation}%
with Ermakov first integral
\begin{equation}
I=\frac{1}{2}\left( x\dot{y}-y\dot{x}\right) ^{2}+\int^{\frac{y}{x}}\left[
\lambda f\left( \lambda \right) -\lambda ^{-3}g\left( \lambda \right) \right]
d\lambda .  \label{KeplerErma.1}
\end{equation}%
The weak Kepler Ermakov system does not admit Lie point symmetries, however
the property of having a first integral prevails. The system of equations (%
\ref{Kepler-Ermakov 01}), (\ref{Kepler-Ermakov 02}) admits the $sl\left(
2,R\right) ~$as Lie point symmetries \cite{LeachK} only for $H\left(
x,y\right) =-\mu ^{2}r^{3}+\frac{h\left( \frac{y}{x}\right) }{x}$~where $\mu$
is either real or pure imaginary number. \ This is the Kepler Ermakov system defined by the equations

\begin{eqnarray}
\ddot{x}+\left( \omega ^{2}(t)-\mu ^{2}\right) x+\frac{1}{r^{3}}h\left(
\frac{y}{x}\right) -\frac{1}{x^{3}}f\left( \frac{y}{x}\right) &=&0
\label{Kepler-Ermakov 1} \\
\ddot{y}+\left( \omega ^{2}(t)-\mu ^{2}\right) y+\frac{1}{r^{3}}\frac{y}{x}%
h\left( \frac{y}{x}\right) -\frac{1}{y^{3}}g\left( \frac{y}{x}\right) &=&0.
\label{Kepler-Ermakov 2}
\end{eqnarray}

It is well known (see \cite{LeachK}) that the oscillator term $\omega
^{2}(t)-\mu ^{2}$ in (\ref{Kepler-Ermakov 1}), (\ref{Kepler-Ermakov 2}) is
removed if one considers new variables $T,X,Y$ defined by the relations:%
\begin{equation}
T=\int \rho ^{-2}dt,X=\rho ^{-1}x~,Y=\rho ^{-1}y  \label{EqnEr.55fa}
\end{equation}%
where $\rho $ is any smooth solution of the time dependent oscillator
equation%
\begin{equation}
\ddot{\rho}+\left( \omega ^{2}(t)-\mu ^{2}\right) \rho =0.
\label{EqnEr.55fb}
\end{equation}

In \cite{LeachK} it is commented that "the effect of $\mu ^{2}$ ($C$ in the
notation of \cite{LeachK}) is to shift the time - dependent frequency
function". However this is true as long as $\omega (t)\neq 0.$ When $\
\omega (t)=0$ one has the autonomous Kepler Ermakov system whose Lie
symmetries span the $sl(2,R)\ $algebra with different representations for $%
\mu =0$ and $\mu \neq 0.$

Before we justify the need for the consideration of the two cases $\mu =0$
and $\mu \neq 0$, we note that by applying the transformation
\begin{equation}
~s=\int v^{-2}dT~,~\bar{x}=v^{-1}X~,~\bar{y}=v^{-1}Y  \label{EqnEr.55fc}
\end{equation}%
where $\nu $ satisfies the Ermakov Pinney equation
\begin{equation}
\frac{d^{2}v}{dT^{2}}+\frac{\mu ^{2}}{v^{3}}=0  \label{EqnEr.55fd}
\end{equation}%
to the transformed equations
\begin{eqnarray}
\frac{d^{2}X}{dT^{2}}+\frac{1}{R^{3}}h\left( \frac{Y}{X}\right) -\frac{1}{%
X^{3}}f\left( \frac{Y}{X}\right) &=&0  \label{EqnEr.55e} \\
\frac{d^{2}Y}{dT^{2}}+\frac{1}{R^{3}}\frac{Y}{X}h\left( \frac{Y}{X}\right) -%
\frac{1}{Y^{3}}g\left( \frac{Y}{X}\right) &=&0  \label{EqnEr.55ff}
\end{eqnarray}%
we retain the term $\mu ^{2}$ and obtain the autonomous Kepler Ermakov
system of \cite{MoyoL}
\begin{eqnarray}
\ddot{x}-\mu ^{2}x+\frac{1}{r^{3}}h\left( \frac{y}{x}\right) -\frac{1}{x^{3}}%
f\left( \frac{y}{x}\right) &=&0  \label{EqnEr.14} \\
\ddot{y}-\mu ^{2}y+\frac{1}{r^{3}}\frac{y}{x}h\left( \frac{y}{x}\right) -%
\frac{1}{y^{3}}g\left( \frac{y}{x}\right) &=&0.  \label{EqnEr.14a}
\end{eqnarray}

The above transformations show that the consideration of the autonomous
Kepler Ermakov system is not a real restriction.

We discuss now the need for the consideration of the cases $\mu =0$ and $\mu
\neq 0.$ In \cite{Tsam10} we have determined the Lie symmetries of the
autonomous 2d Kepler Ermakov system and we have found two cases. The first
Case I concerns the autonomous Kepler Ermakov system with $\mu =0$ and has
the Lie symmetry vectors

\begin{equation}
\mathbf{X}=\left( \bar{c}_{1}+\bar{c}_{2}2t+\bar{c}_{3}t^{2}\right) \partial
_{t}+\left( \bar{c}_{2}+\bar{c}_{3}t\right) r\partial _{r}\qquad (~\mu =0)
\label{Kepler-Ermakov 1b}
\end{equation}

The second Case II concerns the same system with $\mu \neq 0$ and has the
Lie symmetry vectors%
\begin{equation}
\mathbf{X}=\left( c_{1}+c_{2}\frac{1}{\mu }e^{2\mu t}-c_{3}\frac{1}{\mu }%
e^{-2\mu t}\right) \partial _{t}+\left( c_{2}e^{2\mu t}+c_{3}e^{-2\mu
t}\right) ~r\partial _{r}~~~\ ~(~\mu \neq 0)  \label{Kepler-Ermakov 1a}
\end{equation}%
where~in both cases $r\partial _{R}=x\partial _{x}+y\partial _{y}~$is the
gradient HV of the 2d Euclidian metric. Each set of vectors in (\ref%
{Kepler-Ermakov 1a}), (\ref{Kepler-Ermakov 1b}) is a representation of the $%
sl(2,R)$ algebra and furthermore each set of vectors is constructed from the
vector $\partial _{t}$ and the gradient HV $r\partial _{r}$ of the Euclidian
two dimensional space $E^{2}$.

The essence of the difference\ between the two representations is best seen
in the corresponding first integrals. If a Kepler Ermakov system is
Hamiltonian then the Lie symmetries are also Noether point symmetries
therefore in order to find these integrals we determine the Noether
invariants. The Noether symmetries of the Kepler Ermakov system have been
determined in \cite{Tsam10}. For the convenience of the reader we repeat the
relevant material.

Equations (\ref{EqnEr.14}), (\ref{EqnEr.14a}) follow from the Lagrangian
\cite{LeachK}
\begin{equation}
L=\frac{1}{2}\left( \dot{r}^{2}+r^{2}\dot{\theta}^{2}\right) -\frac{\mu ^{2}%
}{2}r^{2}-\frac{C\left( \theta \right) }{2r^{2}}  \label{Kepler-Ermakov 3}
\end{equation}%
where $C(\theta )=c+\sec ^{2}\theta f(\tan \theta )+\csc ^{2}\theta g(\tan
\theta )$ provided the functions $f,g$ satisfy the constraint:%
\begin{equation}
\sin ^{2}\theta f^{\prime }\left( \tan \theta \right) +\cos ^{2}\theta
~g^{\prime }\left( \tan \theta \right) =0.  \label{Kepler-Ermakov 4}
\end{equation}%
The Ermakov invariant in this case is \cite{LeachK}.
\begin{equation}
J=r^{4}\dot{\theta}^{2}+2C\left( \theta \right) .  \label{EqnEr.18a}
\end{equation}

Because the system is autonomous the first Noether integral is the
Hamiltonian \cite{Tsam10}
\begin{equation}
E=\frac{1}{2}\left( \dot{r}^{2}+r^{2}\dot{\theta}^{2}\right) +\frac{1}{2}\mu
^{2}r^{2}+\frac{1}{r^{2}}F\left( \theta \right)
\end{equation}%
In addition to the Hamiltonian there exist two additional time dependent
Noether integrals as follows:\newline
$\mu =0$%
\begin{eqnarray}
I_{1} &=&2tE-r\dot{r}~ \\
I_{2} &=&t^{2}E-tr\dot{r}+\frac{1}{2}r^{2}  \label{Kepler-Ermakov 6}
\end{eqnarray}%
$\mu \neq 0$
\begin{eqnarray}
I_{1}^{\prime } &=&\left( \frac{1}{\mu }E-r\dot{r}+\mu r^{2}\right) e^{2\mu
t}  \label{Kepler-Ermakov 7} \\
I_{2}^{\prime } &=&\left( \frac{1}{\mu }E+r\dot{r}+\mu r^{2}\right) e^{-2\mu
t}.  \label{Kepler-Ermakov 8}
\end{eqnarray}

We note that the Noether integrals corresponding to the representation (\ref%
{Kepler-Ermakov 1b}) are linear in $t$ whereas the ones corresponding to the
representation (\ref{Kepler-Ermakov 1a}) are exponential. Therefore the
consideration of the cases $\mu =0$ and $\mu \neq 0$ is not spurious
otherwise we loose important information. This latter fact is best seen in
the applications of Noether symmetries to field theories where the main core
of the theory is the Lagrangian. In these cases the potential is given and,
as it has been shown in \cite{Tsam10}, a given potential admits certain
Noether symmetries only; therefore one has to consider all possible cases.
We shall come to this situation in section \ref{The Riemannian Kepler
Ermakov system in cosmology} where it will be found that the potential
selects the representation (\ref{Kepler-Ermakov 1a}).

To complete this section we mention that for a Hamiltonian Kepler Ermakov
system the Ermakov invariant (\ref{EqnEr.18a}) is constructed \cite{MoyoL}
from the Hamiltonian and the Noether invariants (\ref{Kepler-Ermakov 7}),(%
\ref{Kepler-Ermakov 8}) as follows:
\begin{equation*}
J=E^{2}-~I_{1}^{\prime }I_{2}^{\prime }.
\end{equation*}

Finally in \cite{MoyoL} it is shown that the Ermakov invariant is generated
by a dynamical Noether symmetry of the Lagrangian (\ref{Kepler-Ermakov 3}),
a result which is also confirmed in \cite{Haas Goedert 2001}.

\section{Generalizing the autonomous Kepler Ermakov system}

\label{Generalizing the Kepler Ermakov system}

We consider the generalization of the two dimensional autonomous Kepler
Ermakov system \cite{Leach1991, Goedert Haas
1998,Reid2,AthorneC,GovinderL,GovinderL2} using a geometric point of view.
From the results presented so far we have the following:

(i) Equations (\ref{EqnEr.50}), (\ref{EqnEr.51}) which define the Ermakov
system employ coordinates in the Euclidian two dimensional space, therefore
the system is the \emph{Euclidian} Ermakov system.

(ii) The autonomous 2d Euclidian Kepler Ermakov system is defined by
equations (\ref{EqnEr.14}) and (\ref{EqnEr.14a})

(iii) The Lie symmetries of the Kepler Ermakov system span the $sl(2,R)$
algebra. These symmetries are constructed from the vector $\partial_t$ and
the gradient HV of the space $E^{2}$.

(iv) For the autonomous Hamiltonian Kepler Ermakov system the Lie symmetries
reduce to Noether symmetries and the Ermakov invariant follows from a
combination of the resulting three Noether integrals, two of which are time
dependent. Furthermore, the Ermakov invariant is the Noether integral of a
dynamical Noether symmetry.

The above observations imply that we may generalize the Kepler Ermakov
system in two directions:

a. Increase the number of dimensions by defining the $n-$dimensional
Euclidian Kepler Ermakov system and/or

b. Generalize the background Euclidian space to be a Riemannian space and
obtain the Riemannian Kepler Ermakov system.

Concerning the defining characteristics of the Kepler Ermakov system we
distinguish three different properties of reduced generality: The property
of having a first integral; the property of admitting Lie/Noether symmetries
and $sl(2,R)$ invariance and the property of being Hamiltonian and admitting
$sl(2,R)$ invariance via Noether symmetries.

Following Leach \cite{Leach1991} we generalize the autonomous Kepler Ermakov
system to higher dimensions by the requirement:

\begin{center}
\textit{The generalized autonomous (Euclidian) Kepler Ermakov system admits
the }$sl(2,R)$\textit{\ algebra as a Lie symmetry algebra.}
\end{center}

In order to exploit the significance of the $sl(2,R)$ invariance we refer to
two theorems, which relate the Lie and the Noether point symmetries of an
autonomous dynamical system moving in a Riemannian space with the symmetries
of the space (for a detailed statement of these theorems see \cite{Tsam10}).

\begin{theorem}
\label{The general conservative system} The Lie point symmetries of the
equations of motion of an autonomous dynamical system moving under the force
$F^{j}(x^{i})$ in a Riemannian space with metric $g_{ij},$ namely%
\begin{equation}
\ddot{x}^{i}+\Gamma _{jk}^{i}\dot{x}^{j}\dot{x}^{k}=F^{i}  \label{PP.01}
\end{equation}%
are given in terms of the generators of the special projective algebra of
the metric $g_{ij}$.
\end{theorem}

If the equations of motion follow form the standard Lagrangian
\begin{equation}
L\left( x^{j},\dot{x}^{j}\right) =\frac{1}{2}g_{ij}\dot{x}^{i}\dot{x}%
^{j}-V\left( x^{j}\right)  \label{NPC.02}
\end{equation}%
where $V(x^{j})$ is the potential, then the following theorem relates the
Noether point symmetries of the Lagrangian with the symmetries of the metric.

\begin{theorem}
\label{The Noether Theorem} The Noether point symmetries of the Lagrangian (%
\ref{NPC.02}) of an autonomous Hamiltonian system moving in a Riemannian
space with metric $g_{ij}$ are generated from the homothetic algebra of $%
g_{ij}$.
\end{theorem}

\section{The 3d autonomous Euclidian Kepler Ermakov system}

\label{The 3d Euclidian Kepler Ermakov system}

The generalization of the autonomous Euclidian Kepler Ermakov system using $%
sl(2,R)$ invariance of Lie symmetries has been done in \cite%
{Leach1991,GovinderL,GovinderL2}. In this section, using of the results of
\cite{DynamicalS}, we give the generalization of the autonomous Euclidian
Hamiltonian Kepler Ermakov system to three dimensions by demanding $sl(2,R)$
invariance with respect to Noether symmetries. The reason for attempting
this generalization is that it leads to the potentials for which the
corresponding extended systems are Liouville integrable.\textbf{\ }%
Furthermore indicates the path to the $n-$dimensional Riemannian Kepler
Ermakov system.

Depending on the value $\mu \neq 0$ or $\mu =0$ we consider the three
dimensional Hamiltonian Kepler Ermakov systems of type I and type II.

\subsection{The 3d autonomous Hamiltonian Kepler Ermakov system of type I $(%
\protect\mu \neq 0)$}

For $\mu \neq 0$ the admitted Noether symmetries are required to be (see (%
\ref{Kepler-Ermakov 1a})):%
\begin{equation}
X^{1}=\partial _{t},~X_{\pm }=\frac{1}{\mu }e^{\pm 2\mu t}\partial _{t}\pm
e^{\pm 2\mu t}R\partial _{R}.  \label{AEP.05.0a}
\end{equation}%
From table 6, line 2 for $T=\frac{1}{\mu }e^{\pm 2\mu t}$ of \cite%
{DynamicalS} we find that for these vectors the potential~is $V\left( R,\phi
,\theta \right) =-\frac{~\mu ^{2}}{2}R^{2}+\frac{1}{R^{2}}f\left( \theta
,\phi \right) ~$hence the Lagrangian%
\begin{equation}
L=\frac{1}{2}\left( \dot{R}^{2}+R^{2}\dot{\phi}^{2}+R^{2}\sin ^{2}\phi ~\dot{%
\theta}^{2}\right) +\frac{~\mu ^{2}}{2}R^{2}-\frac{1}{R^{2}}f\left( \theta
,\phi \right) .  \label{AEP.00}
\end{equation}%
The equations of motion, that is, the equations defining the generalized
dynamical system are
\begin{eqnarray}
\ddot{R}-R\dot{\phi}^{2}-R\sin ^{2}\phi ~\dot{\theta}^{2}-\mu ^{2}R-\frac{2}{%
R^{3}}f &=&0  \label{AEP.01a} \\
\ddot{\phi}+\frac{2}{R}\dot{R}\dot{\phi}-\sin \phi \cos \phi ~\dot{\theta}%
^{2}+\frac{1}{R^{4}}f_{,\phi } &=&0  \label{AEP.01b} \\
\ddot{\theta}+\frac{2}{R}\dot{R}\dot{\theta}+\cot \phi ~\dot{\theta}\dot{\phi%
}+\frac{1}{R^{4}\sin ^{2}\phi }f_{,\theta } &=&0.  \label{AEP.01c}
\end{eqnarray}%
The Noether integrals corresponding to the Noether vectors are%
\begin{eqnarray}
E &=&\frac{1}{2}\left( \dot{R}^{2}+R^{2}\dot{\phi}^{2}+R^{2}\sin ^{2}\phi ~%
\dot{\theta}^{2}\right) -\frac{~\mu ^{2}}{2}R^{2}+\frac{1}{R^{2}}f\left(
\theta ,\phi \right)  \label{AEP.02} \\
I_{+} &=&\frac{1}{\mu }e^{2\mu t}E-e^{2~\mu t}R\dot{R}+\mu e^{2\mu t}R^{2}
\label{AEP.03} \\
I_{-} &=&\frac{1}{\mu }e^{-2\mu t}E+e^{-2\mu t}R\dot{R}+\mu e^{-2\mu t}R^{2}
\label{AEP.04}
\end{eqnarray}%
where $E$ is the Hamiltonian. The Noether integrals $I_{\pm }$ are time
dependent. Following \cite{MoyoL} we define the time independent combined
first integral%
\begin{equation}
J=E^{2}-~I_{+}I_{-}=R^{4}\dot{\phi}^{2}+R^{4}\sin ^{2}\phi ~\dot{\theta}%
^{2}+2f\left( \theta ,\phi \right) .  \label{AEP.05}
\end{equation}%
Using (\ref{AEP.05}) the equation of motion (\ref{AEP.01a})\ becomes
\begin{equation}
\ddot{R}-\mu ^{2}R=\frac{J}{R^{3}}  \label{AEP.05b}
\end{equation}%
which is the autonomous Ermakov- Pinney equation \cite{Pinney}. Therefore $J$
is the Ermakov invariant \cite{Leach1991}.

An alternative way to construct the Ermakov invariant (\ref{AEP.05}) is to
use dynamical Noether symmetries \cite{Kalotas}. Indeed one can show that
the Lagrangian (\ref{AEP.00}) admits the dynamical Noether symmetry $%
X_{D}=K_{j}^{i}\dot{x}^{j}\partial _{i}$ where $K_{ij}$ is a Killing tensor
of the second rank whose non-vanishing components are $K_{\phi \phi
}=R^{4}~,~K_{\theta \theta }=R^{4}\sin ^{2}\phi .~$The dynamical Noether
symmetry vector is~$X_{D}=R^{2}\left( \dot{\phi}\partial _{\phi }+\dot{\theta%
}\partial _{\theta }\right) ~$with gauge function $2f\left( \theta ,\phi
\right) $.

\subsection{The 3d autonomous Hamiltonian Kepler Ermakov system of type II $(%
\protect\mu =0)$}

For $\mu =0$ the Noether symmetries are required to be (see (\ref%
{Kepler-Ermakov 1b})) \cite{Leach1991}:%
\begin{equation}
X^{1}=\partial _{t}~,~X^{2}=2t\partial _{t}~+R\partial
_{R}~,~X^{3}=t^{2}\partial _{t}~+tR\partial _{R}.  \label{AEP.05c}
\end{equation}%
From table 5, line 3 for $T(t)=1$ and table 6, line 1 for $T(t)=t$ of \cite%
{DynamicalS} we find that the potential $V\left( R,\phi ,\theta \right) =%
\frac{1}{R^{2}}f\left( \theta ,\phi \right) $ hence the Lagrangian%
\begin{equation}
L^{\prime }=\frac{1}{2}\left( \dot{R}^{2}+R^{2}\dot{\phi}^{2}+R^{2}\sin
^{2}\phi \dot{\theta}^{2}\right) -\frac{1}{R^{2}}f\left( \theta ,\phi
\right) .  \label{AEP.06}
\end{equation}%
The equations of motion are (\ref{AEP.01a}) - (\ref{AEP.01c}) with $\mu =0$%
.~The Noether invariants of the Lagrangian (\ref{AEP.06}) are%
\begin{eqnarray}
E &=&\frac{1}{2}\left( \dot{R}^{2}+R^{2}\dot{\phi}^{2}+R^{2}\sin ^{2}\phi
\dot{\theta}^{2}\right) +\frac{1}{R^{2}}f\left( \theta ,\phi \right)
\label{AEP.07} \\
I_{1} &=&2tE^{\prime }-R\dot{R}  \label{AEP.08} \\
I_{2} &=&t^{2}E^{\prime }-tR\dot{R}+\frac{1}{2}R^{2}.  \label{AEP.09}
\end{eqnarray}

We note that the time dependent first integrals $I_{1,2}$ are linear in $t$
whereas the corresponding integrals $I_{\pm }$ of the case $\mu \neq 0$ are
exponential. From $I_{1,2}$ we define the time independent first integral $%
J=4I_{2}E^{\prime }-I_{1}^{2}$ which is calculated to be
\begin{equation}
J=R^{4}\dot{\phi}^{2}+R^{4}\sin ^{2}\phi ~\dot{\theta}^{2}+2f\left( \theta
,\phi \right) .  \label{AEP.10}
\end{equation}%
Using (\ref{AEP.10}) the equation of motion for $R\left( t\right) $ becomes~$%
\ddot{R}-\frac{J^{\prime }}{R^{3}}=0$\ which is the one dimensional
Ermakov-Pinney equation, hence $J^{\prime }$ is the Ermakov invariant \cite%
{Leach1991}. As it was the case with the three dimensional Hamiltonian
Kepler Ermakov system of type I, the Lagrangian (\ref{AEP.06}) admits the
dynamical Noether symmetry $X_{D}=R^{2}\left( \dot{\phi}\partial _{\phi }+%
\dot{\theta}\partial _{\theta }\right) $ whose\ integral is the (\ref{AEP.10}%
).

\subsection{Integrability of the 3d autonomous Hamiltonian Euclidian Kepler
Ermakov system}

The 3d autonomous Hamiltonian Euclidian Kepler Ermakov systems need three
independent first integrals in involution in order to be Liouville
integrable. As we have shown each system has the two Noether integrals $%
(E,J) $ , hence we need one more Noether symmetry. Such a symmetry exists
only for special forms of the arbitrary function $f\left( \theta ,\phi
\right) $. From tables 5, 6 and 9-11 of \cite{DynamicalS} we find that extra
Noether symmetries are possible only\footnote{%
The linear combination of an element of $so(3)$ with a translation does not
give a potential, hence an additional Noether symmetry.} for linear
combinations of translations (i.e. vectors of the form $\sum%
\limits_{A=1}^{3}a^{A}\partial _{A}$ where $a^{A}$ are constants) and/or
rotations (i.e. elements of $so(3))$.

\subsubsection{ Noether symmetries generated from the translation group}

We determine the functions $f\left( \theta ,\phi \right) $ for which the 3d
autonomous Hamiltonian Euclidian Kepler Ermakov systems admit extra Noether
symmetries for linear combinations of the translation group.

\noindent \underline{The Lagrangian (\ref{AEP.00})}\newline

In Cartesian coordinates the Lagrangian (\ref{AEP.00}) is%
\begin{equation}
L\left( x^{j},\dot{x}^{j}\right) =\frac{1}{2}\left( \dot{x}^{2}+\dot{y}^{2}+%
\dot{z}^{2}\right) +\frac{\mu ^{2}}{2}\left( x^{2}+y^{2}+z^{2}\right) -\frac{%
1}{x^{2}}f_{I}\left( \frac{y}{x},\frac{z}{x}\right)  \label{LIEP.01}
\end{equation}%
where $f_{I}=\left( 1+\frac{y^{2}}{x^{2}}+\frac{z^{2}}{x^{2}}\right) ^{-1}.$
From table 6, line 1 of \cite{DynamicalS} with $m=-\mu ^{2}~,~p=0$ we find
that the Lagrangian (\ref{LIEP.01}) admits Noether symmetries which are
produced from a linear combination of translations, $\ $if the function $%
f_{I}\left( \frac{y}{x},\frac{z}{x}\right) $ has the form%
\begin{equation}
f_{I}\left( \frac{y}{x},\frac{z}{x}\right) =\frac{1}{\left( 1-\frac{a}{b}%
\frac{y}{x}\right) ^{2}}F\left( \frac{b\frac{z}{x}-c\frac{y}{x}}{\left( 1-%
\frac{a}{b}\frac{y}{x}\right) }\right) .  \label{LIEP.02}
\end{equation}

In this case the Lagrangian (\ref{LIEP.01}) admits at least the following
two extra Noether symmetries (see table 9, line 1 of \cite{DynamicalS})%
\begin{equation}
X_{\pm }=e^{\pm \mu t}{\sum\limits_{A=1}^{3}}a^{A}\partial _{A}
\label{LIEP.03a}
\end{equation}%
with corresponding Noether integrals%
\begin{equation}
I_{\pm }=e^{\pm \mu t}\left( {\sum\limits_{A=1}^{3}}a^{A}\dot{x}_{A}\right)
\mp \mu e^{\pm \mu t}\left( \sum\limits_{A=1}^{3}a^{A}x_{A}\right) .
\label{LIEP.04}
\end{equation}%
We note that the first integrals $I_{\pm }$ are time dependent; however the
first integral
\begin{equation}
J_{2}=I_{+}I_{-}=\left( a\dot{x}+b\dot{y}+c\dot{z}\right) ^{2}+\mu
^{2}\left( ax+by+cz\right) ^{2}  \label{LIEP.05}
\end{equation}%
is time independent. As it was the case with the Ermakov invariant (\ref%
{AEP.05}) the integral $J_{2}$ is possible to be constructed directly from
the dynamical Noether symmetry $X_{D}^{\prime }=K_{\left( 2\right) .j}^{i}%
\dot{x}^{i}\partial _{i}$, where $K_{(2)ij}$ is a Killing tensor of the
second rank \cite{Kalotas,Crampin}, with non-vanishing components
\begin{eqnarray*}
K_{11} &=&a^{2}~,~K_{22}=b^{2}~,~K_{33}=c^{2} \\
K_{\left( 12\right) } &=&2ab~,~K_{\left( 13\right) }=2ac~,~K_{\left(
23\right) }=2bc
\end{eqnarray*}%
so that the dynamical symmetry vector is
\begin{equation}
X_{D}^{\prime }=\left( a^{2}+ab+ac\right) \dot{x}\partial _{x}+\left(
b^{2}+ab+bc\right) \dot{y}\partial _{y}+\left( c^{2}+ac+bc\right) \dot{z}%
\partial _{z}.
\end{equation}%
The Ermakov invariant $J$ (see (\ref{AEP.05}) ) in Cartesian coordinates is
\begin{equation}
J=2E\left( x^{2}+y^{2}+z^{2}\right) -\left( x\dot{x}+y\dot{y}+z\dot{z}%
\right) ^{2}.  \label{LIEP.05b}
\end{equation}

The first integrals $J,J_{2}$ are not in involution. Using the Poisson
brackets we construct new first integrals and at some stage one of them will
be in involution. These new first integrals can also be constructed form
corresponding dynamical Noether symmetries.

An example of a\ known Lagrangian of the form (\ref{LIEP.01}) is the three
body Calogero-Moser Lagrangian \cite{Wojcie,HaasJPA,Ranada}
\begin{equation}
L=\frac{1}{2}\left( \dot{x}^{2}+\dot{y}^{2}+\dot{z}^{2}\right) -\frac{\mu
^{2}}{2}\left( x^{2}+y^{2}+z^{2}\right) -\frac{1}{\left( x-y\right) ^{2}}-%
\frac{1}{\left( x-z\right) ^{2}}-\frac{1}{\left( y-z\right) ^{2}}.
\end{equation}%
The extra Noether symmetries of this Lagrangian are produced by the vector (%
\ref{LIEP.03a})$~$for $a^{A}=\left( 1,1,1\right) .$

\noindent \underline{The Lagrangian (\ref{AEP.06})}\newline

In Cartesian coordinates the Lagrangian (\ref{AEP.06}) is%
\begin{equation}
L\left( x^{j},\dot{x}^{j}\right) =\frac{1}{2}\left( \dot{x}^{2}+\dot{y}^{2}+%
\dot{z}^{2}\right) -\frac{1}{x^{2}}f_{II}\left( \frac{y}{x},\frac{z}{x}%
\right) .  \label{LIEP.07}
\end{equation}%
According to table 10, line 1 and table 11, line 1 of \cite{DynamicalS}
(with $m=0~,~p=0$) the Lagrangian (\ref{LIEP.07}) admits extra Noether point
symmetries for a linear combination of translations if the function $f$ is
of the form (\ref{LIEP.02}). In this case the corresponding Noether
integrals are%
\begin{equation}
I_{1}^{\prime }={\sum\limits_{A=1}^{3}}a^{A}\dot{x}_{A}~,~I_{2}^{\prime }=t{%
\sum\limits_{A=1}^{3}}a^{A}\dot{x}_{A}-{\sum\limits_{A=1}^{3}}a^{A}x_{A}.
\label{LIEP.10}
\end{equation}

Example of such a Lagrangian is the Calogero-Moser Lagrangian \cite{Wojcie}
(without the oscillator term)
\begin{equation}
L=\frac{1}{2}\left( \dot{x}^{2}+\dot{y}^{2}+\dot{z}^{2}\right) -\frac{1}{%
\left( x-y\right) ^{2}}-\frac{1}{\left( x-z\right) ^{2}}-\frac{1}{\left(
y-z\right) ^{2}}.  \label{LIEP.11}
\end{equation}

For the Lagrangian (\ref{LIEP.11}) we have the first integrals $%
E,J,I_{1}^{\prime },I_{2}^{\prime }$. From the integrals $J,I_{1}^{\prime }$
we construct the first integral $\Phi =\left\{ I_{1}^{\prime },\left\{
J,I_{1}^{\prime }\right\} \right\} $ . It is easy to show that the integrals
$E,I_{1}^{\prime },\Phi $ are in involution hence the dynamical system is
Liouville integrable. We remark that the first integrals $E,J,I_{1}^{\prime
},I_{2}^{\prime }$ can also be computed by making use of the Lax pair tensor
\cite{Ranada}.

\subsubsection{Noether symmetries generated from $so\left( 3\right) $}

The elements of $so\left( 3\right) $ in spherical coordinates are the three
vectors $CK^{1,2,3}~\ $%
\begin{equation}
CK^{1}=\sin \theta \partial _{\phi }+\cos \theta \cot \phi \partial _{\theta
},~CK^{2}=\cos \theta \partial _{\phi }-\sin \theta \cot \phi \partial
_{\theta },~CK^{3}=\partial _{\theta }  \label{CCS0.1}
\end{equation}%
which are also the KVs of the Euclidian sphere.

In this case the symmetry condition {c.f. (eq.(56) of \cite{Tsam10})} becomes%
\begin{equation}
L_{CK}\left[ \frac{1}{R^{2}}f\left( \theta ,\phi \right) \right] +p=0~
\label{CCS0.2}
\end{equation}%
or, equivalently%
\begin{equation}
\frac{1}{R^{2}}\left( R^{2}g_{ij}CK^{i}f^{,j}\right) +p=0\Rightarrow
g_{ij}CK_{\left( 1,2,3\right) }^{i}f^{,j}+p=0  \label{CCS0}
\end{equation}%
where $g_{ij}$ is the metric of the Euclidian sphere, that is%
\begin{equation}
ds^{2}=d\phi ^{2}+\mathrm{\sin }^{2}\phi ~d\theta ^{2}.  \label{CCS}
\end{equation}

We infer that the problem of determining the extra Noether point symmetries
of Lagrangian (\ref{AEP.00}) generated from elements of the $so\left(
3\right) $ is equivalent to the determination of the Noether point
symmetries for motion on the 2d sphere, a problem which has been answered in
\cite{Tsam10}.

It is easy to show that the integrals of table 7 of \cite{DynamicalS} are in
involution with the Hamiltonian and the Ermakov invariant, hence the system
is Liouville integrable via Noether symmetries.

The above results are extended to the case in which the system moves on the
hyperbolic sphere that is, it has Lagrangian
\begin{equation}
L=\frac{1}{2}\left( \dot{R}^{2}+R^{2}\dot{\phi}^{2}+R^{2}\sinh ^{2}\phi ~%
\dot{\theta}^{2}\right) +\frac{\mu ^{2}}{2}R^{2}-\frac{1}{R^{2}}g\left(
\theta ,\phi \right) .  \label{EP.SO3}
\end{equation}

We reach at the following conclusion.

\begin{proposition}
The three dimensional autonomous Hamiltonian Kepler Ermakov system with
Lagrangian (\ref{AEP.00}) is Liouville Integrable via Noether point
symmetries which are generated from a linear combination of the three
elements of the~$so\left( 3\right) $ algebra, if and only if the equivalent
dynamical system in the fundamental hyperquadrics\footnote{%
See \cite{Eisenhart} for definitions.} of the three dimensional flat space
is integrable.
\end{proposition}

We note that it is possible a three dimensional autonomous Kepler Ermakov
system to admit more Noether symmetries, which are due to the rotation group
and the translation group (but not to a linear combination of elements from
the two groups). For example the 3d Kepler Ermakov system with Lagrangian
\cite{Damianou2004}%
\begin{equation}
L=\frac{1}{2}\left( \dot{x}^{2}+\dot{y}^{2}+\dot{z}^{2}\right) -\frac{1}{%
x^{2}\left( 1-\frac{y}{x}-\frac{z}{x}\right) ^{2}}
\end{equation}%
has the following extra Noether point symmetries (in addition to the
elements of $sl\left( 2,R\right) $)%
\begin{eqnarray*}
Y^{1} &=&\partial _{x}+\partial _{y},~Y^{2}=\partial _{x}+\partial _{z} \\
Y^{3} &=&t\left( \partial _{x}+\partial _{y}\right) ,~Y^{4}=t\left( \partial
_{x}+\partial _{z}\right) \\
Y^{5} &=&\left( y-z\right) \partial _{x}-\left( x+z\right) \partial
_{y}+\left( x+y\right) \partial _{z}.
\end{eqnarray*}%
The vectors $Y^{1,2}$ and $Y^{3,4}$ follow from (\ref{LIEP.03a}) for $%
a_{1}^{A}=\left( 1,1,0\right) ~$\ and $a_{2}^{A}=\left( 1,0,1\right) \ \ \ $%
respectively whereas $Y^{5}$ is a linear combination of the three elements
of $so\left( 3\right) $. The \ Noether integrals of the Noether symmetries $%
Y^{1-5}$ are respectively
\begin{eqnarray}
I_{Y_{1}} &=&\dot{x}+\dot{y}~,~I_{Y_{2}}=\dot{x}+\dot{z} \\
I_{Y_{3}} &=&t\left( \dot{x}+\dot{y}\right) -\left( x+y\right)
~,~I_{Y_{4}}=t\left( \dot{x}+\dot{z}~\right) -\left( x+z\right) \\
I_{Y_{5}} &=&\left( y-z\right) \dot{x}-\left( x+z\right) \dot{y}+\left(
x+y\right) \dot{z}.
\end{eqnarray}%
It is clear that in order to extend the Kepler Ermakov system to higher
dimensions one needs to have the type of results of \cite{DynamicalS};
therefore the remark made in \cite{Leach1991} that the `notion is easily
generalized to higher dimensions' has to be understood as referring to the
general scenario and not to the actual work.

\section{The autonomous Riemannian Kepler Ermakov system}

\label{The Riemannian Kepler Ermakov system}

As it has been noted in section \ref{ErmakovSys} the Kepler Ermakov systems
considered so far in the literature are the Newtonian Kepler Ermakov
systems. In this section we make a drastic step forward and introduce the
autonomous Riemannian Kepler Ermakov system of dimension $n.$ The
generalization we consider is based on the following definition

\begin{definition}
The $n-$dimensional autonomous Riemannian Kepler Ermakov system is an
autonomous dynamical system which:\newline
a. It is defined on a Riemannian space which admits a gradient HV\newline
b. Admits a first integral, which we name the Riemannian Ermakov first
integral and it is characterized by the requirement that the corresponding
equation of motion takes the form of the Ermakov Pinney equation.\newline
c. It is invariant at least under the $sl(2,R)$ algebra, which is generated
by the vector $\partial_t$ and the gradient HV\ of the space.
\end{definition}

There are two types of $n-$dimensional autonomous Riemannian Kepler Ermakov
systems. The ones which are not Hamiltonian and admit the $sl(2,R)$ algebra
as Lie symmetries and the ones which are Hamiltonian and admit the $sl(2,R)$
algebra as Noether symmetries.

\subsection{The non Hamiltonian autonomous Riemannian Kepler Ermakov system}

\label{The non conservative Riemannian Kepler Ermakov system}

Consider a $n-$dimensional Riemannian space which admits a gradient HV. It
is well known that the metric of this space can always be written in the
form \cite{Tupper1989,Tupper1990}
\begin{equation}
ds^{2}=du^{2}+u^{2}h_{AB}dy^{A}dy^{B}  \label{NCGE.01}
\end{equation}%
where the Latin capital indices $A,B,..$ take the values $1,\ldots,n-1~$and $%
h_{AB}=h_{AB}\left( y^{C}\right) $ is the generic $n-1$ metric. The gradient
HV of the metric is the vector $H^{i}=u\partial _{u}$ ($\psi =1$)$~$%
generated from the function $H=\frac{1}{2}u^{2}$. We note the relation%
\begin{equation}
h_{DA}\Gamma _{BC}^{A}=\frac{1}{2}h_{DB,C}  \label{NCGE.01a}
\end{equation}%
where $\Gamma _{BC}^{A}$ are the connection coefficients of the $n-1$ metric
$h_{AB}.$ In that space we consider a particle moving under the action of
the force $F^{i}=F^{u}(u,y^{C})\frac{\partial }{\partial u}+$ $F^{A}(u,y^{C})%
\frac{\partial }{\partial y^{A}}.$ The equations of motion $\frac{Dx^{i}}{Dt}%
=F^{i}$ when projected along the direction of $u$ and in the $n-1$ space
give the equations
\begin{eqnarray}
u^{\prime \prime }-uh_{AB}y^{\prime A}y^{\prime B} &=&F^{u}  \label{NCGE.02}
\\
y^{\prime \prime A}+\frac{2}{u}u^{\prime }y^{\prime A}+\Gamma
_{BC}^{A}y^{\prime B}y^{\prime C} &=&F^{A}  \label{NCGE.03}
\end{eqnarray}%
where $u^{\prime }=\frac{du}{ds}$ and $s~$is an affine parameter.

Because the system is autonomous admits the Lie point symmetry $\partial
_{t} $. Using the vector $\partial _{t}$ and the gradient HV $%
H^{i}=u\partial _{u} $ we construct two representations of $sl(2,R)$ by
means of the sets of vectors (see (\ref{AEP.05.0a}),(\ref{AEP.05c}))%
\begin{equation}
\partial _{s},~2s\partial _{s}+u\partial _{u},~s^{2}\partial _{t}+su\partial
_{u}\qquad \text{ when }\mu =0  \label{NCGE.04}
\end{equation}%
\begin{equation}
\partial _{s},~\frac{1}{\mu }e^{\pm 2\mu s}\partial _{s}\pm e^{\pm 2\mu
s}u\partial _{u}\qquad ~~~\text{ when }\mu \neq 0  \label{NCGE.05}
\end{equation}%
and require that the vectors in each set will be Lie symmetries of the
system of equations (\ref{NCGE.02}),(\ref{NCGE.03}). In Appendix A we show
that the requirement of the invariance of the force under both
representations (\ref{NCGE.04}), (\ref{NCGE.05}) of $sl(2,R)$ demands that
the force is of the form%
\begin{equation}
F^{i}=\left( \mu ^{2}u+\frac{1}{u^{3}}G^{u}\left( y^{C}\right) \right)
\partial _{u}+\frac{1}{u^{4}}G^{A}\left( y^{C}\right) \partial _{A}.
\label{NCGE.05a}
\end{equation}%
Replacing $F^{i}$ in the system of equations (\ref{NCGE.02}),(\ref{NCGE.03})
we find
\begin{eqnarray}
u^{\prime \prime }-uh_{AB}y^{\prime A}y^{\prime B} &=&\mu ^{2}u+\frac{1}{%
u^{3}}G^{u}  \label{NCGE.06} \\
y^{\prime \prime A}+\frac{2}{u}u^{\prime }y^{\prime A}+\Gamma
_{BC}^{A}y^{\prime B}y^{\prime C} &=&\frac{1}{u^{4}}G^{A}.  \label{NCGE.07}
\end{eqnarray}%
\ Multiplying the second equation with $2u^{4}h_{DA}y^{\prime D}$ and using (%
\ref{NCGE.01a}) we have%
\begin{equation}
u^{4}\frac{d}{ds}\left( h_{DB}y^{\prime D}y^{\prime B}\right)
+4u^{3}h_{DA}u^{\prime }y^{\prime A}y^{\prime D}=2G_{D}y^{\prime D}
\end{equation}%
from which follows%
\begin{equation}
\frac{d}{ds}\left( u^{4}h_{DB}y^{\prime D}y^{\prime B}\right)
=2G_{D}y^{\prime D}.
\end{equation}

The rhs is a perfect differential if~$G_{D}=-\Sigma _{,D}~$where $\Sigma
(y^{A})$ is a differentiable function. If this is the case we find the first
integral
\begin{equation}
J=u^{4}h_{DB}y^{\prime D}y^{\prime B}+2\Sigma \left( y^{C}\right).
\label{NCGE.010}
\end{equation}%
We note that $J$ involves the arbitrary metric $h_{AB}$ and the function $%
\Sigma (y^{A}$). Furthermore equations (\ref{NCGE.06}), (\ref{NCGE.07}%
)~become
\begin{eqnarray}
u^{\prime \prime }-uh_{AB}y^{\prime A}y^{\prime B} &=&\mu ^{2}u+\frac{1}{%
u^{3}}G^{u}\left( y^{C}\right)  \label{NCGE.011} \\
y^{\prime \prime A}+\frac{2}{u}u^{\prime }y^{\prime A}+\Gamma
_{BC}^{A}y^{\prime B}y^{\prime C} &=&-\frac{1}{u^{4}}h^{AB}\Sigma \left(
y^{C}\right) _{,B}~.  \label{NCGE.012}
\end{eqnarray}%
These are the equations defining the $n-$dimensional autonomous Riemannian
Kepler Ermakov system.

Using the first integral (\ref{NCGE.010}) the equation of motion (\ref%
{NCGE.011}) is written as follows
\begin{equation}
u^{\prime \prime }=\mu ^{2}u+\frac{\bar{G}\left( y^{C}\right) }{u^{3}}
\end{equation}%
where $\bar{G}=J+G^{u}\left( y^{C}\right) -2\Sigma \left( y^{C}\right) .$
This is the Ermakov-Pinney equation hence we identify (\ref{NCGE.010}) as
the Riemannian Ermakov integral of the autonomous Riemannian Kepler Ermakov
system.

\subsection{The autonomous Hamiltonian Riemannian Kepler Ermakov system}

\label{The Hamiltonian Kepler Ermakov system in a n-dimensional Riemannian
space}

In this section we assume that the force is derived from the potential $%
V\left( u,y^{C}\right) .$ Then the equations of motion follow from the
Lagrangian%
\begin{equation}
L=\frac{1}{2}\left( u^{\prime 2}+u^{2}h_{AB}y^{\prime A}y^{\prime B}\right)
-V\left( u,y^{C}\right) .  \label{GERS.02}
\end{equation}%
The Hamiltonian is%
\begin{equation}
E=\frac{1}{2}\left( u^{\prime 2}+u^{2}h_{AB}y^{\prime A}y^{\prime B}\right)
+V\left( u,y^{C}\right) .  \label{GERS.03}
\end{equation}%
The equations of motion are%
\begin{eqnarray}
u^{\prime \prime }-uh_{AB}y^{\prime A}y^{\prime B}+V_{,u} &=&0
\label{GERS.03a} \\
y^{\prime \prime A}+\frac{2}{u}u^{\prime }y^{\prime A}+\Gamma
_{BC}^{A}y^{\prime B}y^{\prime C}+\frac{1}{u^{2}}h^{AB}V_{,B} &=&0.
\label{GERS.03b}
\end{eqnarray}%
The demand that Lagrangian (\ref{GERS.02}) admits Noether point symmetries
which are generated from the gradient HV leads to the following cases (see
theorem \ref{The Noether Theorem} and for details see \cite{Tsam10})

Case A: The Lagrangian (\ref{GERS.02}) admits the Noether point symmetries (%
\ref{NCGE.04}) if the potential is of the form
\begin{equation}
V\left( u,y^{C}\right) =\frac{1}{u^{2}}V\left( y^{C}\right) .  \label{GERS.5}
\end{equation}%
The Noether integrals of these Noether point symmetries are
\begin{eqnarray}
E_{A} &=&\frac{1}{2}\left( u^{\prime 2}+u^{2}h_{AB}y^{\prime A}y^{\prime
B}\right) +\frac{1}{u^{2}}V\left( y^{C}\right)  \label{GERSN.1} \\
I_{1} &=&2sE-uu^{\prime }  \label{GERSN.2} \\
I_{2} &=&s^{2}E-suu^{\prime }+\frac{1}{2}u^{2}  \label{GERSN.3}
\end{eqnarray}%
where $E_{A}$ is the $\ $Hamiltonian.

Case B: The Lagrangian (\ref{GERS.02}) admits the Noether point symmetries (%
\ref{NCGE.05}) if the potential is of the form
\begin{equation}
V\left( u,y^{c}\right) =-\frac{\mu ^{2}}{2}u^{2}+\frac{1}{u^{2}}V^{\prime
}\left( y^{C}\right) .  \label{GERS.6}
\end{equation}%
The Noether integrals of these Noether point symmetries are
\begin{eqnarray}
E_{B} &=&\frac{1}{2}\left( u^{\prime 2}+u^{2}h_{AB}y^{\prime A}y^{\prime
B}\right) -\frac{\mu ^{2}}{2}u^{2}+\frac{1}{u^{2}}V^{\prime }\left(
y^{C}\right)  \label{GERSN.4} \\
I_{+} &=&\frac{1}{\mu }e^{2\mu s}E-e^{2\mu s}uu^{\prime }+\mu e^{2\mu s}u^{2}
\label{GERSN.5} \\
I_{-} &=&\frac{1}{\mu }e^{-2\mu s}E+e^{-2\mu s}uu^{\prime }+\mu e^{-2\mu
s}u^{2}  \label{GERSN.6}
\end{eqnarray}%
where $E_{B}$ is the $\ $Hamiltonian.

Using the Noether integrals we construct the Riemannian Ermakov invariant $%
J_{G}$, which is common for both Case A and Case B, as follows%
\begin{equation}
J_{G}=u^{4}h_{DB}y^{\prime D}y^{\prime C}+2V^{\prime }\left( y^{C}\right) .
\label{GERSN.7}
\end{equation}%
This coincides with the invariant first integral defined in (\ref{NCGE.010}%
). We note that with the use of the first integral (\ref{GERSN.7}) the
Hamiltonians (\ref{GERSN.1}) and (\ref{GERSN.4}) take the form
\begin{equation}
E=\frac{1}{2}u^{\prime 2}-\frac{\mu ^{2}}{2}u^{2}+\frac{J}{2u^{2}}
\end{equation}%
which is the Hamiltonian for the Ermakov Pinney equation.

As it was the case with the Euclidian case of section \ref{ErmakovSys}, it
can be shown that the Riemannian Ermakov invariant (\ref{GERSN.7}) is due to
a dynamical Noether symmetry\cite{Kalotas}.

We collect the results in the following proposition.

\begin{proposition}
\label{prop ghv}In a Riemannian space with metric $g_{ij}$ which admits a
gradient HV, the equations of motion of a Hamiltonian system moving under
the action of the potential $(\mu \epsilon
\mathbb{C}
)$
\begin{equation}
V\left( u,y^{c}\right) =-\frac{\mu ^{2}}{2}u^{2}+\frac{1}{u^{2}}V^{\prime
}\left( y^{C}\right)  \label{GERSN.8}
\end{equation}%
admit the $sl(2,R)\ $invariance and also an invariant first integral, the
Riemannian Ermakov invariant. This latter quantity is also possible to be
identified as the Noether integral of a dynamical Noether symmetry.
\end{proposition}

Without going into details we state the following more general result.

\begin{proposition}
\label{prop gkv}Consider a $n-$ dimensional Riemannian space with an $r$--
decomposable metric \ which in the Cartesian coordinates $x_{1},\ldots,x_{r}$
has the general form%
\begin{equation}
ds^{2}=p\eta _{\Sigma \Lambda }dz^{\Sigma }dz^{\Lambda
}+h_{ij}dx^{i}dx^{j}~\ ,~i,j=r+1,\ldots,n~,~\Sigma =1,\ldots,r
\label{GERSN.10}
\end{equation}%
where $\eta _{\Sigma \Lambda }$ is a flat non-degenerate metric (of
arbitrary signature). If there exists a potential so that the vectors $%
e^{\pm \mu s}{\sum\limits_{M}}a^{M}\partial _{^{M}}$ , are Noether
symmetries, where $a_{M}$ are constants, with Noether integrals $~$%
\begin{equation}
I_{\pm }=e^{\pm \mu s}\sum\limits_{M}a^{M}z_{M}^{\prime }\mp \mu e^{\pm \mu
s}\sum\limits_{M}a^{M}z_{M}  \label{GERSN.11}
\end{equation}%
the combined first integral $I=I_{+}I_{-}$ is time independent and it is the
result of a dynamical Noether symmetry.
\end{proposition}

In the remaining sections we consider applications of the autonomous
Riemannian Kepler Ermakov system in General Relativity and in Cosmology.

\section{The autonomous Riemannian Kepler Ermakov system in General
Relativity}

\label{The RKEGR}

\subsection{The autonomous Riemannian Kepler Ermakov system of four degrees
of freedom in a FRW spacetime}

Consider the spatially flat FRW spacetime with metric
\begin{equation}
ds^{2}=du^{2}-u^{2}\left( dx^{2}+dy^{2}+dz^{2}\right) .  \label{Apl.1}
\end{equation}%
This metric admits the gradient HV $u\partial _{u}$ and six non gradient KVs
\cite{Maartens} which are the KVs of $E^{3}$.

We consider the autonomous Riemannian Kepler Ermakov system defined by the
Lagrangian (see (\ref{GERSN.8}) ) ($\mu\in \mathbb{C}$ )
\begin{equation}
L=\frac{1}{2}\left( u^{\prime 2}-u^{2}\left( x^{\prime 2}+y^{\prime
2}+z^{\prime 2}\right) \right) +\frac{\mu ^{2}}{2}u^{2}-\frac{1}{u^{2}}%
V\left( x,y,z\right) .  \label{Apl.2}
\end{equation}

The Euler Lagrange equations are%
\begin{eqnarray}
u^{\prime \prime }+u\left( x^{\prime 2}+y^{\prime 2}+z^{\prime 2}\right)
-\mu ^{2}u-\frac{2V\left( x,y,z\right) }{u^{3}} &=&0 \\
x^{\sigma \prime \prime }+\frac{2}{u}u^{\prime }x^{\sigma \prime }-\frac{%
V^{,\sigma }\left( x,y,z\right) }{u^{4}} &=&0
\end{eqnarray}%
where $\sigma =1,2,3$. The Lagrangian (\ref{Apl.2}) has the form of the
Lagrangian (\ref{GERS.02}) for potential $V\left( u,y^{C}\right) =-\frac{\mu
^{2}}{2}u^{2}+\frac{1}{u^{2}}V\left( x,y,z\right) $ hence according to
proposition \ref{prop ghv} possesses $sl(2,R)$ with Noether symmetries and
for \emph{both} representations (\ref{NCGE.04}) and (\ref{NCGE.05}). The two
time independent invariants are the Hamiltonian and the Riemannian Ermakov
invariant (proposition \ref{prop ghv})
\begin{eqnarray}
E &=&\frac{1}{2}\left( u^{\prime 2}-u^{2}\left( x^{\prime 2}+y^{\prime
2}+z^{\prime 2}\right) \right) -\frac{\mu ^{2}}{2}u^{2}+\frac{1}{u^{2}}%
V\left( x,y,z\right) \\
J_{G_{4}} &=&u^{4}\left( x^{\prime 2}+y^{\prime 2}+z^{\prime 2}\right)
+2V\left( x,y,z\right) .
\end{eqnarray}

We \ remark that if we had considered the representation\ (\ref{NCGE.04})
only (that is we had set $\mu =0)$ then we would have lost all information
concerning the system defined for $\mu \neq 0!$ We emphasize that in
application to Physics the major datum is the Lagrangian and not in the
equations of motion, therefore one should not make mathematical assumptions
which restrict the physical generality.

To assure Liouville integrability we need one more Noether symmetry whose
Noether integral is in involution with $E,J_{G_{4}}.$ This is possible for
certain forms of the potential $V\left( x,y,z\right) .$ Using the general
results of \cite{DynamicalS} where all 3d potentials are given which admit
extra Noether symmetries we find the following result.

\begin{proposition}
\label{FRW_Flat}The Lagrangian (\ref{Apl.2}) admits an extra Noether
symmetry if and only if the potential $V\left( x,y,z\right) $ has the form
given in table 5, lines: 1,2. table 8, lines :1-4 and table 9, lines:1-4 of
\cite{DynamicalS}.
\end{proposition}

For example if $V\left( x,y,z\right) =\left( x^{2}+y^{2}+z^{2}\right) ^{n}$
then the system admits three extra Noether symmetries which are the elements
of $so\left( 3\right) $. If $V\left( x,y,z\right) =V_{0}$ then the system
admits six extra Noether symmetries (the KVs of the three dimensional
Euclidian space).

\subsection{The autonomous Riemannian Kepler Ermakov system of three degrees
of freedom in a three dimensional spacetime}

Consider the three dimensional Lorenzian metric
\begin{equation}
ds^{2}=du^{2}-u^{2}\left( dx^{2}+dy^{2}\right)  \label{AGE.00}
\end{equation}%
which admits the gradient HV $u\partial _{u}$ and the three KVs of the
Euclidian metric $E^{2}.$ In that space consider the Lagrangian
\begin{equation}
L^{\prime }=\frac{1}{2}\left( u^{\prime 2}-u^{2}\left( x^{\prime
2}+y^{\prime 2}\right) \right) +\frac{\mu ^{2}}{2}u^{2}-\frac{1}{u^{2}}%
V\left( x,y\right) .  \label{AGE.01}
\end{equation}%
According to proposition \ref{prop ghv} this Lagrangian admits as Noether
symmetries the elements of $sl\left( 2,R\right) .~$Then from proposition \ref%
{prop ghv}\ we have that the Noether\ invariants of these symmetries are%
\begin{eqnarray}
E &=&\frac{1}{2}\left( u^{\prime 2}-u^{2}\left( x^{\prime 2}+y^{\prime
2}\right) \right) -\frac{\mu ^{2}}{2}u^{2}+\frac{1}{u^{2}}V\left( x,y\right)
\\
J_{G_{3}} &=&u^{4}\left( x^{\prime 2}+y^{\prime 2}\right) +2V\left(
x,y\right) .
\end{eqnarray}

The requirement that the Lagrangian admits an additional Noether symmetry
leads to the condition $L_{KV}V\left( x,y\right) +p=0$ therefore in that
case we have a 2d potential and we can use the results of \cite{Tsam10}. If
we demand the new Noether integral to be time independent~$\left( p=0\right)
$ then the potential $V\left( x,y\right) $ and the new Noether integrals are
given in Table 1 of Appendix B.

Lagrangians with kinetic term $T_{K}=\frac{1}{2}\left( u^{\prime
2}-u^{2}\left( x^{\prime 2}+y^{\prime 2}\right) \right) ~$appear in
cosmological models. In the following section we discuss such applications.

\section{The autonomous Riemannian Kepler Ermakov system in cosmology}

\label{The Riemannian Kepler Ermakov system in cosmology} In a locally
rotational symmetric (LRS) spacetime we consider two cosmological models for
dark energy, a scalar field cosmology and an $f(R)$ cosmology.

\subsection{The case of scalar field cosmology}

\label{scalar field cosmology}

Consider the Class A LRS spacetime
\begin{equation}
ds^{2}=-N^{2}\left( t\right) dt^{2}+a^{2}\left( t\right) e^{-2\beta \left(
t\right) }dx+a^{2}\left( t\right) e^{\beta \left( t\right) }\left(
dy^{2}+dz^{2}\right)  \label{LRS.01}
\end{equation}%
which is assumed to contain a scalar field with exponential potential $%
V\left( \phi \right) =V_{0}e^{-c\phi }~,~c\neq \sqrt{6k}$ and a perfect
fluid with a stiff equation of state $p=\rho $, where $p$ is the pressure
and $\rho $ is the energy density of the fluid. The conservation equation
for the matter density gives
\begin{equation}
\dot{\rho}+6\rho \frac{\dot{a}}{a}=0\rightarrow \rho =\frac{\rho _{0}}{a^{6}}%
.  \label{LRS.02}
\end{equation}

Einstein field equations for the comoving observers $u^{a}=\frac{1}{N\left(
t\right) }\partial _{t}~,~u^{a}u_{a}=-1$~follow from the autonomous
Lagrangian \cite{Rayn,KotsakisL}
\begin{equation}
L=-3\frac{a}{N}\dot{a}^{2}+\frac{3}{4}\frac{a^{3}}{N}\dot{\beta}+\frac{k}{2}%
\frac{a^{3}}{N}\dot{\phi}^{2}-Nka^{3}e^{-c\phi }-N\frac{\rho _{0}}{a^{3}}.
\label{LRS.03}
\end{equation}%
We set $N^{2}=e^{c\phi }$ and the Lagrangian becomes
\begin{equation}
L=(-3\dot{a}^{2}+\frac{3}{4}a^{2}\dot{\beta}^{2}+\frac{k}{2}a^{2}\dot{\phi}%
^{2})ae^{-\frac{c}{2}\phi }-ka^{3}V_{0}e^{-\frac{c}{2}\phi }-\frac{\rho _{0}%
}{a^{3}e^{\frac{c}{2}\phi }}.  \label{LRS.04}
\end{equation}%
The Hamiltonian is%
\begin{equation}
E=(-3\dot{a}^{2}+\frac{3}{4}a^{2}\dot{\beta}^{2}+\frac{k}{2}a^{2}\dot{\phi}%
^{2})ae^{-\frac{c}{2}\phi }+ka^{3}V_{0}e^{-\frac{c}{2}\phi }+\frac{\rho _{0}%
}{a^{3}e^{\frac{c}{2}\phi }}=0
\end{equation}%
If we consider the transformation%
\begin{equation}
a^{3}=e^{x+y}~,~\phi =\frac{1}{3}\sqrt{\frac{6}{k}}\left( x-y\right)
\label{LRS.07}
\end{equation}%
where
\begin{equation}
x=\frac{1}{1-\bar{c}}\ln \left( \frac{\left\vert 1-\bar{c}\right\vert }{%
\sqrt{2}}ue^{z}\right) ~,~y=\frac{1}{1+\bar{c}}\ln \left( \frac{1+\bar{c}}{%
\sqrt{2}}ue^{-z}\right) ~,~\bar{c}=\frac{c}{\sqrt{6k}}\neq 1  \label{LRS.08}
\end{equation}%
the Lagrangian (\ref{LRS.04}) becomes
\begin{equation}
L=-\frac{2}{3}\dot{u}^{2}+u^{2}\left( \frac{2}{3}\dot{z}^{2}-\frac{3}{8kV_{0}%
}\dot{\beta}^{2}\right) -\frac{\mu ^{2}}{2}u^{2}+\frac{kV_{0}\rho _{0}}{\mu
^{2}}\frac{1}{u^{2}}~\ ,~\mu ^{2}=kV_{0}\left( 1-\bar{c}^{2}\right) \neq 0.
\label{LRS.09}
\end{equation}

We consider a 2d Riemannian space with metric defined by the kinematic terms
of the Lagrangian, that is:%
\begin{equation}
ds^{2}=\left( -6da^{2}+\frac{3}{2}a^{2}d\beta ^{2}+ka^{2}d\phi ^{2}\right)
ae^{-\frac{c}{2}\phi }  \label{LRS.05}
\end{equation}

We show easily that this metric admits the gradient HV $H^{i}=\frac{4}{%
6k-c^{2}}\left( ka\partial _{a}+c\partial _{\phi }\right) ~,~$with gradient
function~$H=\frac{8k\varepsilon }{c^{2}-6k}ae^{-\frac{c}{2}\phi }.~\ $
Therefore the Lagrangian (\ref{LRS.04}) defines an autonomous Hamiltonian
Riemannian Kepler Ermakov system with potential ($\mu \neq 0$)
\begin{equation}
V(u,y^{A})=-\frac{1}{2}\mu ^{2}u^{2}+\frac{kV_{0}\rho _{0}}{\mu ^{2}}\frac{1%
}{u^{2}}.  \label{LRS.09a}
\end{equation}
Because $\mu \neq 0$ this Lagrangian admits $sl(2,R)$ invariance only for
the representation (\ref{NCGE.04}) (an additional result which shows the
necessity for the consideration of the cases $\mu =0$ and $\mu \neq 0!).$

Using proposition \ref{prop ghv} we write the Ermakov invariant
\begin{equation}
J=u^{4}\left( \frac{2}{3}\dot{z}^{2}+\frac{3}{8kV_{0}}\dot{\beta}^{2}\right)
+\frac{kV_{0}\rho _{0}}{\mu ^{2}}\frac{1}{u^{2}}.  \label{LRS.11}
\end{equation}%
The second invariant is the Hamiltonian $E~$
\begin{equation}
E=-\frac{2}{3}\dot{u}^{2}+u^{2}\left( \frac{2}{3}\dot{z}^{2}+\frac{3}{8kV_{0}%
}\dot{\beta}^{2}\right) +\frac{\mu ^{2}}{2}u^{2}-\frac{kV_{0}\rho _{0}}{\mu
^{2}}\frac{1}{u^{2}}.  \label{LRS.12}
\end{equation}

From table 1 line 6 of \ \cite{DynamicalS} \ we find that the Lagrangian
admits three more Noether symmetries%
\begin{equation}
\partial _{\beta },~\partial _{z}~,~z\partial _{\beta }-\beta \partial _{z}
\label{LRS.13}
\end{equation}%
with corresponding integrals%
\begin{equation}
I_{1}=u^{2}\dot{\beta}~,~I_{2}=u^{2}\dot{z}~,~I_{3}=u^{2}\left( \frac{3}{%
8kV_{0}}z\dot{\beta}-\frac{2}{3}\beta \dot{z}\right) .  \label{LRS.14}
\end{equation}%
It is easy to show that three of the integrals are in involution, therefore
the system is Liouville integrable.

\subsection{The case of $f\left( R\right) $ - cosmology}

\label{frgravity}

Consider the modified Einstein-Hilbert action
\begin{equation}
S=\int d^{4}x\sqrt{-g}f\left( R\right)  \label{MLRS.01}
\end{equation}%
where $f(R)$ is a smooth function of the curvature scalar $R.$ The resulting
field equations for this action in the metric variational approach are \cite%
{Sotiriou}
\begin{equation}
f^{\prime }R_{ab}-\frac{1}{2}fg_{ab}+g_{ab}\square f^{\prime
}-f_{;ab}^{\prime }=0  \label{MLRS.02}
\end{equation}%
where $f^{\prime }=\frac{df\left( R\right) }{dR}~$and $f^{\prime \prime
}\neq 0.~$In the LRS spacetime (\ref{LRS.01}) with $N\left( t\right) =1$
these equations for comoving observers are the Euler-Lagrange equations of
the Lagrangian
\begin{equation}
L=\left( 6af^{\prime }\dot{a}^{2}+6a^{2}f^{\prime \prime }\dot{a}\dot{R}-%
\frac{3}{2}f^{\prime }a^{3}\dot{\beta}^{2}\right) +a^{3}\left( f^{\prime
}R-f\right) .  \label{MLRS.03}
\end{equation}%
The Hamiltonian is:%
\begin{equation}
E=\left( 6af^{\prime }\dot{a}^{2}+6a^{2}f^{\prime \prime }\dot{a}\dot{R}-%
\frac{3}{2}f^{\prime }a^{3}\dot{\beta}^{2}\right) -a^{3}\left( f^{\prime
}R-f\right) =0.  \label{MLRS.05}
\end{equation}%
Again we consider the 3d Riemannian space whose metric is defined by the
kinematic part of the Lagrangian (\ref{MLRS.03})
\begin{equation}
ds^{2}=12af^{\prime }da^{2}+12a^{2}f^{\prime \prime }da~dR-3a^{3}f^{\prime
}d\beta ^{2}.  \label{MLRS.04}
\end{equation}%
This metric admits the gradient HV%
\begin{equation}
H^{i}=\frac{1}{2}\left( a\partial _{a}+\frac{f^{\prime }}{f^{\prime \prime }}%
\partial _{R}\right)  \label{MLRS.06}
\end{equation}%
with gradient function $H=3a^{3}f^{\prime }.$

In order to determine the function $f(R)$ we demand the geometric condition
that Lagrangian (\ref{MLRS.03}) admits $s(2,R)$ invariance via Noether
symmetries (see \cite{Tsam10}). Then for each representation (\ref{NCGE.04}%
), (\ref{NCGE.05}) we have a different function $f(R)$ hence a different
physical theory.

The representation (\ref{NCGE.04}) in the present context is:%
\begin{equation}
\partial _{t}~,~2t\partial _{t}+\frac{1}{2}\left( a\partial _{a}+\frac{%
f^{\prime }}{f^{\prime \prime }}\partial _{R}\right) ~,~t^{2}\partial _{t}+%
\frac{t}{2}\left( a\partial _{a}+\frac{f^{\prime }}{f^{\prime \prime }}%
\partial _{R}\right) .  \label{SL.2R1}
\end{equation}%
The Noether conditions become
\begin{equation}
-4a^{3}f^{\prime }R+\frac{7}{2}a^{3}f+p=0.
\end{equation}%
These vectors are Noether symmetries if $p=0$ and
\begin{equation}
f\left( R\right) =R^{\frac{7}{8}}.
\end{equation}%
however power law $f\left( R\right) $ theories are not cosmologically viable
\cite{Amendola2}.

The second representation (\ref{NCGE.05}) in the present context gives the
vectors%
\begin{equation}
\partial _{t}~,~\frac{1}{\mu }e^{\pm 2\mu t}\partial _{t}\pm \frac{1}{2}%
e^{\pm 2\mu t}\left( a\partial _{a}+\frac{f^{\prime }}{f^{\prime \prime }}%
\partial _{R}\right) .  \label{SL.2R2}
\end{equation}%
The Noether conditions give
\begin{equation}
-4a^{3}f^{\prime }R+\frac{7}{2}a^{3}f+3\mu ^{2}a^{3}f^{\prime }+p=0.
\label{SL.2R2.1}
\end{equation}%
These vectors are Noether symmetries if the constant $p=0$ and \ the
function
\begin{equation}
f\left( R\right) =\left( R-2\Lambda \right) ^{\frac{7}{8}}  \label{SL.2R2.2}
\end{equation}%
where $2\Lambda =3\mu ^{2}.$ This model is the viable $\Lambda _{bc}$%
CDM-like cosmological with $b=1,c=\frac{7}{8}$. \cite{Amendola}.

We note that if we had not considered the latter representation then we
would loose this interesting \footnote{%
The importance of the result is due to the fact that it follows from a
geometric assumption which is beyond and above the physical considerations.
Furthermore the assumption of Noether symmetries provides the Noether
integrals which allow for an analytic solution of the model.} result.

For the function (\ref{SL.2R2.2}) the Lagrangian (\ref{MLRS.03}) becomes for
both cases (if $\Lambda =0$ we have the power-law $f\left( R\right) =R^{%
\frac{7}{8}}$)
\begin{eqnarray}
L &=&\frac{21}{4}a\left( R-2\Lambda \right) ^{-\frac{1}{8}}\dot{a}^{2}-\frac{%
21}{16}a^{2}\left( R-2\Lambda \right) ^{-\frac{9}{8}}\dot{a}\dot{R}  \notag
\\
&-&\frac{21}{8}a^{3}\left( R-2\Lambda \right) ^{-\frac{1}{8}}\dot{\beta}^{2}-%
\frac{a^{3}}{8}\frac{\left( R-16\Lambda \right) }{\left( R-2\Lambda \right)
^{\frac{1}{8}}}.  \label{MLRS.10}
\end{eqnarray}%
Furthermore there exist a coordinate transformation for which the metric (%
\ref{MLRS.04}) is written in the form of (\ref{GERS.02}).

We introduce new variables $u,v,w$ with the relations
\begin{equation}
a=\left( \frac{21}{4}\right) ^{-\frac{1}{3}}\sqrt{ue^{v}}~,~R=2\Lambda +%
\frac{e^{12v}}{u^{4}},~\beta =\sqrt{2}w.
\end{equation}%
In the new variables the Lagrangian (\ref{MLRS.10}) takes the form
\begin{equation}
L=\frac{1}{2}\dot{u}^{2}-\frac{1}{2}u^{2}\left( \dot{v}^{2}+\dot{w}%
^{2}\right) +\frac{\mu ^{2}}{2}u^{2}-\frac{1}{42}\frac{e^{12v}}{u^{2}}.
\label{MLRS.11}
\end{equation}%
The Hamiltonian (\ref{MLRS.05}) in the new coordinates is%
\begin{equation}
E=\frac{1}{2}\dot{u}^{2}-\frac{1}{2}u^{2}\left( \dot{v}^{2}+\dot{w}%
^{2}\right) -\frac{\mu ^{2}}{2}u^{2}+\frac{1}{42}\frac{e^{12v}}{u^{2}}.
\end{equation}%
The Lagrangian (\ref{MLRS.11}) defines a Hamiltonian Riemannian Kepler
Ermakov system with potential%
\begin{equation*}
V\left( u,v\right) =-\frac{\mu ^{2}}{2}u^{2}+\frac{1}{42}\frac{e^{12v}}{u^{2}%
}
\end{equation*}%
from which follows the potential $V(v)=\frac{1}{42}e^{12v}.~\ $ In addition
to the Hamiltonian the dynamical system admits the Riemannian Ermakov
invariant%
\begin{equation}
J_{f}=u^{4}\left( \dot{v}^{2}+\dot{w}^{2}\right) +\frac{1}{21}e^{12v}.
\end{equation}

The Lagrangian (\ref{MLRS.11}) admits the extra Noether point symmetry $%
\partial _{w}$ with Noether integral $I_{w}=u^{2}\dot{w}$ ~(see Table 1).
The three integrals $E,I_{w}~$and $J_{f}$ are in involution and independent,
therefore the system is integrable.

\section{Conclusion}

\label{Conclusion}

In this work we have considered the generalization of the autonomous Kepler
Ermakov dynamical system in the spirit of Leach \cite{Leach1991}, that is
using invariance wrt the $sl(2,R)$ Lie and Noether algebra. We have
generalized the autonomous Newtonian Hamiltonian Kepler Ermakov system to
three dimensions using Noether rather than Lie symmetries and have
determined all such systems which are Liouville integrable via Noether
symmetries. We introduced the autonomous Riemannian Kepler Ermakov system in
a Riemannian space which admits a gradient HV. This system is the
generalization of the autonomous Euclidian Kepler Ermakov system and opens
new fields of applications for the autonomous Kepler Ermakov system,
especially in relativistic Physics. Indeed we have determined the autonomous
Riemannian Kepler Ermakov system in a spatially flat FRW\ spacetime which
admits a gradient HV. As a further application we have considered two types
of cosmological models which are described by the autonomous Riemannian
Kepler Ermakov system, i.e. scalar field cosmology with exponential
potential and $f(R)-$ gravity in an LRS spacetime.

\subsection*{Acknowledgement}

We would like to thank Professor Leach P. G. L. and the referees for
critical comments and suggestions. This work has been partially supported
from ELKE (grant 1112) of the University of Athens.

\section*{ Appendix A: Proof of the force in equations (\protect\ref{NCGE.02}%
), (\protect\ref{NCGE.03}).}

\label{apen1}

We require that the force admits two Lie symmetries which are due to the
gradient HV $H=u\partial _{u}~$(if we require the force to be invariant
under three Lie symmetries which are due to the gradient HV then it is
reduced to the isotropic oscillator). From theorem \ref{The general
conservative system} (for details see \cite{Tsam10})\ we have the following
cases.

(I) Case $\mu =0$\newline
In this case the Lie symmetries are
\begin{equation*}
\partial _{s},~2s\partial _{s}+u\partial _{u},~s^{2}\partial _{s}+su\partial
_{u}.
\end{equation*}%
The condition which the force must satisfy is%
\begin{equation*}
L_{H}F^{i}+dF^{i}=0.
\end{equation*}%
Replacing components we find the equations%
\begin{eqnarray*}
\left( \frac{\partial }{\partial u}F^{u}\right) u+\left( d-1\right) F^{u}
&=&0 \\
\left( \frac{\partial }{\partial u}F^{A}\right) u+dF^{A} &=&0
\end{eqnarray*}%
from which follows
\begin{equation*}
F^{u}=\frac{1}{u^{\left( d-1\right) }}F^{u}~,~F^{A}=\frac{1}{u^{d}}F^{A}.
\end{equation*}%
Because the HV\ is gradient, condition A2 of theorem \ref{The general
conservative system} applies and gives the condition
\begin{equation*}
L_{H}F^{i}+4F^{i}+a_{1}H^{i}=0
\end{equation*}%
from which follows $a_{1}=0$ and \ $d=4$. Therefore
\begin{equation*}
F^{u}=\frac{1}{u^{3}}G^{u}\left( y^{C}\right) ~,~F^{A}=\frac{1}{u^{4}}%
G^{A}\left( y^{C}\right) .
\end{equation*}

(II) Case $\mu \neq 0$\newline
In this case the Lie symmetries are
\begin{equation*}
\partial _{s},~\frac{1}{\mu }e^{\pm 2\mu s}\partial _{s}\pm e^{\pm 2\mu
s}u\partial _{u}
\end{equation*}%
The condition which the force must satisfy is%
\begin{equation*}
L_{H}F^{i}+4F^{i}+a_{1}H^{i}=0
\end{equation*}%
We demand $a_{1}\neq 0$ and obtain the system of equations:%
\begin{eqnarray*}
\left( \frac{\partial }{\partial u}F^{u}\right) u+3F^{u}+a_{1}u &=&0 \\
\left( \frac{\partial }{\partial u}F^{A}\right) u+4F^{A} &=&0
\end{eqnarray*}%
whose solution is
\begin{equation*}
F^{u}=\mu ^{2}u+\frac{1}{u^{3}}G^{u}~,~F^{A}=\frac{1}{u^{4}}G^{A}
\end{equation*}%
where we have set $a_{1}=-4\mu ^{2}$.

\section*{Appendix B: Table 1}

\label{appen2}

\begin{center}
Table 1: Potentials for which the Lagrangian (\ref{AGE.01}) admits extra
Noether point symmetry

\
\begin{tabular}{|l|l|l|}
\hline
\textbf{Noether Symmetry} & $\mathbf{V}\left( x,y\right) $ & \textbf{Noether
Integral} \\ \hline
$\partial _{x}$ & $f\left( y\right) $ & $I_{x}=u^{2}x^{\prime }$ \\ \hline
$\partial _{y}$ & $f\left( x\right) $ & $I_{y}=u^{2}y^{\prime }$ \\ \hline
$y\partial _{x}-x\partial _{y}$ & $f\left( x^{2}+y^{2}\right) $ & $%
I_{xy}=u^{2}\left( yx^{\prime }-xy^{\prime }\right) $ \\ \hline
$\partial _{x}+b\partial _{y}$ & $f\left( y-bx\right) $ & $%
I_{xby}=u^{2}\left( x^{\prime }+by^{\prime }\right) $ \\ \hline
$\left( a+y\right) \partial _{x}+\left( b-x\right) \partial _{y}$ & $f\left(
\frac{1}{2}\left( x^{2}+y^{2}\right) +ay-bx\right) $ & $I_{abxy}=\left(
a+y\right) u^{2}x^{\prime }+\left( b-x\right) u^{2}y^{\prime }$ \\ \hline
$\partial _{x},~\partial _{y}~,~y\partial _{x}-x\partial _{y}$ & $V_{0}$ & $%
I_{x}~,I_{y}~,~I_{xy}$ \\ \hline
\end{tabular}
\end{center}

\end{document}